\documentclass[aps,prb,amsfonts,amssymb,%longbibliography,
twocolumn,amsmath,floatfix,showpacs]{revtex4-1}
\usepackage{amsbsy}
\usepackage{subfigure}
\usepackage[dvips]{color}
\usepackage[dvips]{graphics}
\usepackage{graphicx}% Include figure files
\usepackage{bm}% bold math

\newcommand{\be}{\begin{equation}}
\newcommand{\ee}{\end{equation}}
\newcommand{\bel}{\begin{align}}
\newcommand{\eel}{\end{align}}
\newcommand{\bem}{\begin{multline}}
\newcommand{\eem}{\end{multline}}
\newcommand{\beq}{\begin{equation}}
\newcommand{\eeq}{\end{equation}}
\newcommand{\bea}{\begin{eqnarray}}
\newcommand{\eea}{\end{eqnarray}}

\DeclareMathOperator{\sech}{sech}

\begin{document}

\title{Magnetic penetration depth and vortex structure in anharmonic
superconducting junctions with an interfacial pair breaking}

\author{Yu.\,S.~Barash}

\affiliation{Institute of Solid State Physics, Russian Academy of Sciences,
Chernogolovka, Moscow District, 142432 Russia}

\date{May 23, 2014}

\begin{abstract}
The penetration depth $l_{\text{j}}$ in superconducting junctions is
identified within the Ginzburg-Landau theory as a function of the
interfacial pair breaking, of the magnetic field and of the Josephson
coupling strength. When the interfacial pair breaking goes up,
$l_{\text{j}}$ increases and an applicability of the local Josephson
electrodynamics to junctions with a strong Josephson coupling is
extended. In the junctions with strongly anharmonic current-phase
relations, the magnetic field dependence of $l_{\text{j}}$ is shown
to lead to a significant difference between the weak-field penetration
depth and the characteristic size of the Josephson vortex. For such
junctions a nonmonotonic dependence of $l_{\text{j}}$ and of the
lower critical field on the Josephson coupling constant is found, and
the specific features of spatial profiles of the supercurrent and the
magnetic field in the Josephson vortex are established.
\end{abstract}


\maketitle

\section{Introduction}
\label{sec: intro}

The magnetic self-effects of the Josephson current in wide
superconductor-thin interlayer-superconductor junctions result in
the screening inside the junction interlayer of the magnetic field
applied at the junction edge, and in the formation of Josephson
vortices. Spatial variations of the magnetic field and of the
current density along the junction interface are coupled with
variations of the phase difference and should be determined
jointly.

The corresponding results for standard tunnel junctions with a
sufficiently weak Josephson coupling have been obtained
within the Ginzburg-Landau (GL) theory since the early days of studying the
Josephson effect.~\cite{Ferrell1963,Josephson1965,Scalapino1967,%
Barone1982} The results imply that the Josephson penetration depth
$\lambda_{\text{J}}$ significantly exceeds the London penetration depth
$\lambda_{\text{L}} $, as is commonly observed. The Meissner
screening being significantly stronger than the Josephson one is
consistent with the definition of ``weak superconductivity'',
although it is not generally an integral feature of weak links.

The conventional definition of weak links and, in particular of
tunnel junctions, requires the critical current $j_{\text{c}}$ to be
significantly less than the depairing current $j_{\text{dp}}$ deep
within the superconducting leads. On the other hand, the condition
$\lambda_{\text{J}}\gg\lambda_{\text{L}}$ is equivalent to
$j_{\text{c}}^{1/2}\ll\left(\frac{j_{\text{dp}}}{\kappa}\right)^{1/2}
\!\!\!$, where $\kappa$ is the GL parameter. For junctions involving
strongly type-II superconductors, the relation presented can prove to
be more restrictive for the critical current than the weak-link
requirement $j_{\text{c}}\ll j_{\text{dp}}$. Hence, the standard
results only apply to tunnel junctions satisfying the condition
$j_{\text{c}}^{1/2}\ll\left(\frac{j_{\text{dp}}}{\kappa}
\right)^{1/2}\!\!\!$. In the opposite case $j_{\text{c}}^{1/2}\agt
\left(\frac{j_{\text{dp}}}{\kappa}\right)^{1/2}\!\!\!$,\, $
j_{\text{c}}\ll j_{\text{dp}}$, which ensures the relation $
\lambda_{\text{J}}\alt\lambda_{\text{L}}$ for weak links, the
electrodynamics of tunnel junctions acquires a nonlocal
character.~\cite{AGurevich1992,Silin1992} In a strongly nonlocal
regime $\lambda_{\text{J}}\ll\lambda_{\text{L}}$, the characteristic
scale of an isolated Josephson vortex along the junction plane is
$\frac{\lambda_{\text{J}}^2}{\lambda_{\text{L}}}$, which is
substantially less than the Josephson penetration depth
$\lambda_{\text{J}}$.~\cite{AGurevich1992} Recently the nonlocality
has been experimentally identified in planar junctions with thin
superconducting electrodes~\cite{Krasnov2013}, where the conditions
for observing the nonlocal effects~\cite{Ivanchenko1990,%
Humphreys1993,Mints1997,Kogan2001,Mints2008,Clem2010} are modified,
and monitored more easily as compared with the junctions with thick leads.

A distinctive feature of superconducting junctions considered in this
paper is the presence of an interfacial pair breaking. An intense
interfacial pair breaking can take place, for example, in junctions
involving unconventional superconductors and/or magnetic or normal
metal interlayers. Since in a small transition region weak links are
quite sensitive to local conditions, an interface-induced local
weakening of the superconducting condensate density can have a
profound influence on the whole of the Josephson effect. As the result,
the interplay of the Josephson coupling strength and interfacial
pair activity controls the behavior of the supercurrent.

A weak Josephson coupling leads to the sinusoidal (harmonic)
current-phase relation, whereas a strongly anharmonic supercurrent
emerges at the large values of the coupling constant. In planar junctions
with a strong Josephson coupling and vanishing interfacial pair
activity the critical current $j_{\text{c}}$ becomes comparable with
the depairing current $j_{\text{dp}}$, and the junctions do not
represent weak links.~\cite{Golubov2004} Conversely, the critical
current of the junctions with an intense interfacial pair breaking is
strongly suppressed, as compared to the case of no pair breaking, and
can only be substantially less than $j_{\text{dp}}$, irrespective of
the Josephson coupling strength.~\cite{Barash2012} Thus the
interfacial pair breaking maintains the planar junctions with a
pronounced Josephson coupling as weak links $j_{\text{c}}\ll
j_{\text{dp}}$ with strongly anharmonic current-phase relations.

This paper addresses effects of the interfacial pair breaking and of
the Josephson coupling strength on the magnetic penetration depth
$l_{\text{j}}$ and the Josephson vortex structure in wide planar
junctions involving strongly type-II superconductors. For a fixed
Josephson coupling, the quantity $l_{ \text{j}}$ is shown to go up
with the interfacial pair breaking. This substantially extends an
applicability domain of the condition $l_{\text{j}}\gg
\lambda_{\text{L}}$ and, hence, of the local Josephson
electrodynamics to the junctions with a strong Josephson coupling in
the presence of an intense interfacial pair breaking.

The magnetic field dependence of the penetration depth
is studied below both for harmonic and
anharmonic superconducting junctions. In the junctions with the
harmonic supercurrent described by local Josephson electrodynamics,
the Josephson penetration depth $\lambda_{\text{J}}$ is the only
characteristic scale of the problem. Along with the critical current,
it depends substantially on the strength of the interfacial pair
breaking. Under a weak applied field, $\lambda_{\text{J}} $ exactly
coincides with the penetration depth, while the latter is shown to
depend on the magnetic flux $\Phi$ through the junction and to
approach the value $l_{\text{jv}}=l_{\text{j}}(\frac{\Phi_0}{2})=
\frac{\pi}{2} \lambda_{\text{J}}$ at half of the flux quantum.

While in harmonic junctions a characteristic size of the Josephson
vortex $l_{ \text{jv}}$ (a half of its effective width) is of the
same order as $\lambda_{\text{J}}$, in junctions with strongly
anharmonic current-phase relations the magnetic field dependence
$l_{\text{j}}(\Phi)$ is demonstrated to become pronounced and to
result in a significant difference between $l_{\text{jv}}$ and a
weak-field penetration depth $l_{\text{j}0}$. As a specific feature
of the strongly anharmonic current-phase relation, a nonmonotonic
dependence of the Josephson vortex size $l_{\text{jv}}$ and of the
lower critical field on the Josephson coupling strength, for a fixed
and intense interfacial pair breaking, is identified within the local
Josephson electrodynamics. Finally, the spatial structure of an
isolated Josephson vortex in the junctions with an intense
interfacial pair breaking is studied. In particular, narrow peaks in
the current-phase relation of strongly anharmonic junctions are shown
to transform into narrow peaks in a spatial profile of the supercurrent
density in the vortex.

The paper is organized as follows. The magnetic field dependence of
the penetration depth in harmonic junctions is described in
Sec.~\ref{sec: harmonic}. In Sec.~\ref{sec: anharmonic} the
penetration depth in anharmonic junctions is obtained as a function
of the magnetic field, of the Josephson coupling constant and of the
strength of the interfacial pair breaking. Section~\ref{sec: josscreen}
addresses spatial profiles of the phase difference, of the magnetic
field and of the supercurrent density in an isolated Josephson vortex
in anharmonic junctions. The lower critical field in such junctions
is found in Sec.~\ref{sec: lowerfield}. Section~\ref{sec: discussion}
contains discussions and Sec.~\ref{sec: conclusions} concludes the paper.

\section{$l_{\text{j}}$ in harmonic junctions}
\label{sec: harmonic}

\begin{figure}[t] \centering
\includegraphics[width=1.\columnwidth,clip=true]{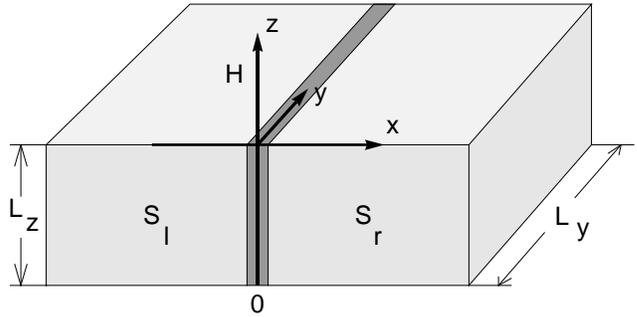}
\caption{Schematic diagram of the junction.} \label{scheme}
\end{figure}

Let the static magnetic field $\bm{H}=H\bm{e}_z$ be applied along the
$z$ axis to a symmetric planar junction involving thick leads made of
strongly type-II superconductors (see Fig.\,\ref{scheme}). A
homogeneous plane rectangular interlayer at $x=0$ is supposed to be
of zero length within the GL approach. The spatially constant widths
$L_y,L_z$ of the junction are considered to significantly exceed the
penetration depths: $L_y,\,L_z\gg l_{\text{j}}, \lambda_{\text{L}}$.
Under such conditions the magnetic field is independent of the $z$
coordinate inside the interlayer and in the superconductors.

The applied field is assumed to be substantially less than the
critical fields of the leads, and to produce a negligibly small
influence on the Josephson current as a function of the phase
difference $j(\chi)$. At the same time the self-field effects,
generated by the current flowing through wide junctions,
interconnect the magnetic field $H[\chi(y)]$, the supercurrent
density $j[\chi(y)]$ and the spatially dependent phase difference
$\chi(y)$, and can have a profound influence on their spatial
distributions.

Within the local Josephson electrodynamics, which presupposes the
condition $l_{\text{j}}\gg \lambda_{\text{L}}$, the spatially
dependent static phase difference $\chi(y)$ in the junctions with a
harmonic current-phase relation $j(\chi)=j_{\text{c}}\sin\chi$
satisfies a well-known one-dimensional sine-Gordon equation
\cite{Ferrell1963,Josephson1965,Scalapino1967,Barone1982}
\begin{equation}
\dfrac{d^2\chi(y)}{dy^2}=\,\dfrac1{\lambda_{\text{J}}^2}\sin\chi(y).
\label{chieq2}
\end{equation}
Here $\lambda_{\text{J}}$ is the Josephson penetration depth
$\lambda_{\text{J}}=\left({c\Phi_0}/{16\pi^2\lambda_{\text{L}}
j_{\text{c}}}\right)^{1/2}$ and $\Phi_0={\pi\hbar c}/{|e|}$ is the
superconductor flux quantum.

The self-consistent results of the GL theory for the Josephson current
$j(\chi)$ in planar junctions \cite{Barash2012,Barash2012_2,[{See
Supplemental Material for details of the derivations and for results
of the GL theory for the Josephson current as a function of $g_\ell$
and $g_\delta$}]suppl}, is being used below.
The order parameters in the two superconducting leads is written as
$f_{1(2)}(x)e^{i\chi_{1(2)}(x)}$, where the moduli $f_{1(2)}(x)$ are
normalized to their values in the bulk in the absence of the
supercurrent. In symmetric junctions $f=f(|x|)$, i.e., $f_{2}(x)=
f_{1}(-x)$, and the boundary conditions for $f$ are
\begin{equation}
\biggl(\!\frac{df}{d\overline{x}}\!\biggr)_{\pm} =
\pm\!\left(g_{\delta}+2g_\ell\sin^2\!\frac{\chi}2\right)f_0,
\label{bcss99}
\end{equation}
where $f_{0}$ is an interface value of $f(x)$, $\overline{x}=x/\xi(T)
$, $\xi(T)$ is the temperature dependent superconductor coherence
length and $\chi$ is the phase difference $\chi=\chi_--\chi_+$.

The coefficient $g_\ell$ in \eqref{bcss99} is the effective
dimensionless Josephson coupling constant, and $g_{\delta}$ is the
effective dimensionless interface parameter. The parameters
$g_\delta$ and $g_\ell$ are the main characteristics of the interface
in the GL theory. They are assumed to be positive and, therefore,
resulting in an interfacial pair breaking in accordance with
\eqref{bcss99}. In the absence of the current, i.e., at $\chi=0$, the
suppression of the order parameter at the interface is described
solely by $g_\delta$. When the supercurrent flows, the Josephson coupling
contributes to the phase dependent suppression of the order parameter
at the interface~\cite{suppl}.

In macroscopic samples of strongly type-II superconductors, the
influence of the interfacial pair breaking on the Meissner effect is
small, in the measure of $\kappa^{-1}\ll1$, and will be disregarded
below. Thus the local penetration depth of the Meissner effect is
considered to be spatially constant, irrespective of the boundary
conditions for the order parameter, and equal to $\lambda_{\text{L}}$
which is related to the bulk condensate density. Contrary to its
negligible influence on the Meissner effect, the interfacial pair
breaking can have a considerable impact both on the critical current
and, in the presence of a pronounced Josephson coupling, on the
current-phase relation. For this reason the standard expression and
estimates for $j(\chi)$, which do not take into account effects of
the interfacial pair breaking and of the Josephson coupling strength,
can fail.

The harmonic
current-phase relation $j(\chi)=j_{\text{c}}\sin\chi$ takes place
under the condition $g_\ell\ll\max\bigl(1,\, g_\delta\bigr)$, which
incorporates not only tunnel junctions, defined as $g_\ell\ll 1$, but
also the junctions with a strong Josephson coupling $1 \alt g_\ell\ll
g_\delta$ in the presence of an intense interfacial pair breaking
\cite{suppl,Barash2012_2}. The junctions satisfying the generalized
condition $g_\ell\ll\max\bigl(1,\,g_\delta\bigr)$ will be called
harmonic junctions. With the corresponding expression for the
critical current of harmonic junctions (see (S10) in \cite{suppl}) and with those for
$\lambda_L$ and $\xi$, the quantity $\lambda_{\text{J}}$ in
\eqref{chieq2} can be written as
\begin{equation}
\lambda_{\text{J}}=\left(\frac{c\Phi_0}{16\pi^2\lambda_{\text{L}}
j_{\text{c}}}\right)^{1/2}=\left(\dfrac{\lambda_{\text{L}}\xi}{g_\ell}
\right)^{1/2}\dfrac1{\sqrt{2+g_\delta^2}-g_\delta}.
\label{lambdaj1}
\end{equation}
Hence, the characteristic length scale $\lambda_{\text{J}}$
substantially depends on the strength of the interfacial pair
breaking $g_\delta$.

In standard tunnel junctions with $g_\delta\ll 1$ one gets
$\lambda_{\text{J}} \gg\lambda_{\text{L}}$, since the parameter
$g_\ell$ is proportional to the junction transparency and  in this
case extremely small. The characteristic length \eqref{lambdaj1}
decreases $\propto g_\ell^{-1/2}$ with increasing the effective
Josephson coupling constant and becomes comparable with
$\lambda_{\text{L}}$ at the characteristic value
$g_\ell^{1/2}\sim\kappa^{-1/2}\ll 1$, which can be still small in the
strongly type-II superconductors. At the same time,
$\lambda_{\text{J}}$ increases with the interfacial pair breaking. In
junctions with an intense pair breaking $g_\delta\gg 1$ the limiting
relation $\lambda_{\text{J}}\approx\left(\frac{\lambda_{\text{L}}
\xi}{g_\ell}\right)^{1/2}g_\delta$ follows from \eqref{lambdaj1}. One
sees that $\lambda_{\text{J}}$ considerably exceeds
$\lambda_{\text{L}}$ under the condition $g_\ell^{1/2}\ll g_\delta
\kappa^{-1/2}$, which allows the strong coupling constant
$g_\ell\agt 1$, provided $g_\delta\gg\kappa^{1/2}$. Thus, in the
presence of an intense interfacial pair breaking the local
electrodynamics can be applied to describing the harmonic junctions
with a large Josephson coupling.

If a strongly nonlocal regime $\lambda_{\text{J}}\ll \lambda_{\text{L}}$
takes place,  one can combine the results of Ref.~\onlinecite{AGurevich1992}
with Eq.~\eqref{lambdaj1}, where the effects of the interfacial pair breaking
are taken into account.
This leads to the following characteristic scale of an isolated Josephson
vortex
\begin{equation}
\frac{\lambda_{\text{J}}^2}{\lambda_{\text{L}}}=\dfrac{\xi}{g_\ell\bigl(
\sqrt{g_\delta^2+2}-g_\delta\bigr)^2}.
\end{equation}

Further on the condition $\lambda_{\text{J}}\gg\lambda_{\text{L}}$ will be
assumed, which ensures an applicability of the local theory. For the
quantitative analysis, let us consider
the junction of Ferrell and Prange~\cite{Ferrell1963,Barone1982},
i.e., a wide junction occupying the halfspace $y>0$, $L_y\to\infty$
under the magnetic field applied at the junction edge $y=0$.  The
magnetic field is assumed to be fully screened far inside the
junction plane ($y\to\infty$), where the supercurrent density also
vanishes. In describing the screening effects, the magnetic flux
$\Phi$ through the junction will be considered not exceeding half
of the flux quantum $|\Phi|\le\frac{\Phi_0}{2}$. The magnetic field
$H_0$ at the junction edge at $\Phi=\frac{\Phi_0}{2}$ is known to be
the highest field, for which a solution with no vortex precursors is
possible, and, therefore, the magnetic field as well as the current
density decay monotonically with increasing the distance $y$ from the
interlayer edge. The screening of such an external field is only
metastable, since it exceeds the lower critical
field.~\cite{Josephson1965,Tinkham1996} At the same time, the spatial
distributions of the quantities $H(y)$, $j(y)$ and $\chi(y)$,
controlled by the screening effect at $\Phi= \frac{\Phi_0}{2}$,
coincide with their spatial profiles in the half of an isolated
Josephson vortex involving single flux quantum $\Phi_0$. Hence, when
$\Phi$ is equal to half of a flux quantum, the penetration
depth $l_{\text{j}}(\frac{\Phi_0}{2})$ represents a characteristic
size $l_{\text{jv}}$ of the vortex, a half of its effective width
along the $y$ axis. One also notes, that the magnetic field in
the center of the vortex, produced by the vortex Josephson current,
coincides with the magnetic field $H_0$ at the junction edge at
$\Phi=\frac{\Phi_0}{2}$.

As a weak applied field $\Phi\ll\Phi_0$ induces only a small
supercurrent in the junction ($|\sin\chi|\ll 1$), one can consider
small phase differences and linearize the sine function in
Eq.~\eqref{chieq2}. This results in a simple exponentially decaying
solution of \eqref{chieq2}: $\chi=\chi_0\exp(-y/\lambda_{\text{J}})$,
\,$H(y)=-\left[\Phi_0\chi_0\bigl/(4\pi\lambda_{\text{L}}
\lambda_{\text{J}})\right]\exp(-y/\lambda_{\text{J}})$. The latter
expression signifies that the quantity \eqref{lambdaj1} coincides
with the weak-field penetration depth exactly:\, $l_{\text{j}0} =
\lambda_{\text{J}}$.~\cite{Barone1982,Tinkham1996} With the
increasing magnetic flux through the junction, the linearized description
fails and one should use the solution of Eq.~\eqref{chieq2} found in
Ref.~\onlinecite{Ferrell1963}. For the magnetic field at $x=0$ inside
the junction $y>0$, with a maximum at the junction edge $y=0$, one
has $H(y)=\mp\Phi_0\bigl/\left[2\pi\lambda_{\text{L}}
\lambda_{\text{J}}\cosh\bigl((y+y_0)\bigl/\lambda_{\text{J}}\bigr)
\right]$, $y_0\ge0$, and the phase difference is $\chi(y)=\pm2
\arcsin\sech\bigl((y+y_0)\bigl/\lambda_{\text{J}}\bigr)$.

Since the spatial profile $H(y)$ of the magnetic field in the
junction interlayer ($x=0$) can substantially differ from the
exponential one, the equality
\begin{equation}
\int\limits_{0}^{+\infty}H(y)dy=l_{\text{j}} H(0)
\label{deflambdaj1}
\end{equation}
will be put to use for a quantitative description of the junction
penetration depth $l_{\text{j}}$. Equation~\eqref{deflambdaj1} is in
agreement with the standard definition of magnetic penetration depths
in various other circumstances.~\cite{LandauECM1984,Tinkham1996} Here
$H(0)$ is the magnetic field at the junction edge $y=0$, and
\eqref{deflambdaj1} defines a characteristic size of an adjacent
region, where the magnetic field as well as the d.c. supercurrent are
confined within the junction.

Substituting the solution for $H(y)$ in \eqref{deflambdaj1} and
taking the integral, one gets $l_{\text{j}}$ as a function of $y_0$.
Since $y_0$ and $\Phi$ are implicitly related to each other in
accordance with the condition $\Phi=2\lambda_{\text{L}} l_{\text{j}}
H(0)$, one obtains eventually the dependence of the Josephson
penetration depth on the magnetic flux through the junction
\begin{equation}
l_{\text{j}}^{-1}(\Phi)=\lambda_{\text{J}}^{-1}
\dfrac{\sin\left(\pi\Phi\bigl/\Phi_0\right)}{\pi\Phi\bigl/\Phi_0},
\quad |\Phi|\le\dfrac12\Phi_0,
\label{lambdaPhi1}
\end{equation}
and the relation
\begin{equation}
H(0)=\dfrac{\Phi_0}{2\pi\lambda_{\text{J}}\lambda_{\text{L}}}
\sin\left(\dfrac{\pi\Phi}{\Phi_0}\right).
\end{equation}
Thus, $l_{\text{j}}(\Phi)$ goes up with the increase of the
magnetic flux within the given limits. While $l_{\text{j}}(\Phi)
\approx l_{\text{j}0} =\lambda_{\text{J}}$ for $\pi|\Phi|\ll\Phi_0$,
one gets $l_{\text{j}}(\frac{\Phi_0}2) \equiv l_{\text{jv}}=\frac{\pi
}{2}\lambda_{\text{J}}$ when half of the flux quantum pierces the
junction. Here both the weak-field penetration depth $l_{\text{j}0}$
and the characteristic size of the Josephson vortex $l_{\text{jv}}$
are associated with one and the same length scale $\lambda_{\text{J}}
$. The difference between them, though quantitatively noticeable, is
not significant.

\section{$l_{\text{j}}$ in junctions with anharmonic current-phase relations}
\label{sec: anharmonic}

In anharmonic junctions the equation for a spatially dependent phase
difference takes the form [cf. \eqref{chieq2}]
\begin{equation}
\dfrac{d^2\chi[(y)]}{dy^2}-\,\dfrac{16\pi^2\lambda_{\text{L}}}{c\Phi_0}
j[\chi(y)]=0,
\label{chieq1}
\end{equation}
and the defining relation for the penetration depth
$l_{\text{j}}(\Phi)$ as a function of the magnetic flux is \cite{suppl}
\begin{equation}
l_{\text{j}}^{-1}(\Phi)=\left[\dfrac{8 \lambda_{\text{L}} \Phi_0}{c\Phi^2}\!\!\!
\int\limits_{0}^{\frac{2\pi\Phi}{\Phi_0}}j(\chi)d\chi\right]^{1/2}\!\!\!\! .
\label{deflambdaj2}
\end{equation}
Here $j(\chi)$, the Josephson current density in the absence of the
magnetic field, is assumed to be an odd function of $\chi$.

Equation~\eqref{deflambdaj2} describes the junction penetration depth and
its magnetic flux dependence, assuming the current-phase relation of
the junction to be known. Therefore, making use of the results of the
GL theory for the anharmonic phase dependence $j(\chi)$,
allows one to obtain from \eqref{deflambdaj2} the quantity
$l_{\text{j}}(\Phi)$. Substituting  $j=j_{\text{c}}\sin\chi$ in
\eqref{deflambdaj2}, one easily reproduces Eq.~\eqref{lambdaPhi1} for
harmonic junctions.

The results for harmonic junctions remain applicable to the
anharmonic case under sufficiently weak applied magnetic fields, when
a spatially dependent current density is small enough throughout the
junction plane allowing the linearization of the current-phase
relation: $j\approx j'_0\chi$,\, $j'_0=\left(\frac{dj(\chi)}{d\chi}
\right)_{\chi=0}$. Then the integration of the current density in
\eqref{deflambdaj2} results in the penetration depth $l_{\text{j}0}=
\left(\frac{c\Phi_0}{16\pi^2\lambda_{\text{L}} j'_0}\right)^{1/2}$.
Though for anharmonic junctions $j'_0$, in general, is not the critical
current, it is so for the harmonic ones. With this proviso, the
weak-field penetration depth $l_{\text{j}0}$ coincides with
$\lambda_{\text{J}}$.

An analytical expression for $l_{\text{j}}$ can be obtained for
arbitrary values of the Josephson coupling $g_{\ell}$ in the regime
of a pronounced interfacial pair breaking $g_{\delta}^2\gg 1$, which
guarantees that the planar junctions are the weak links. Using the
corresponding current-phase relation of the GL theory, one gets from
\eqref{deflambdaj2} \cite{suppl}
\begin{equation}
\l_{\text{j}}(\Phi,g_\ell,g_\delta)=\dfrac{2
\left({\pi|\Phi|}\bigl/{\Phi_0}\right)
\sqrt{\left(g_{\delta}+g_{\ell}\right)
\lambda_{\text{L}}\xi}}{\ln^{1/2}\left[1+
\dfrac{4g_{\ell}\left(g_{\delta}+g_{\ell}\right)}{g_{\delta}^2}
\sin^2\dfrac{\pi\Phi}{\Phi_0}\right]}.
\label{ljPhi2}
\end{equation}

Under the condition $4g_{\ell}\left(g_{\delta}+g_{\ell}\right)\sin^2
\frac{\pi\Phi}{\Phi_0}\ll g_{\delta}^2$, which is satisfied in the
weak-field and/or in the tunneling limits, Eq.~\eqref{ljPhi2} reduces
to \eqref{lambdaPhi1} with $\lambda_{\text{J}}$ defined in
\eqref{lambdaj1} and taken at $g_{\delta}^2\gg 1$.

The junction penetration depth \eqref{ljPhi2}, as a function of the
magnetic flux, monotonically increases from $l_{\text{j}0}=g_{\delta}
\left(\lambda_{\text{L}}\xi\bigl/ g_\ell\right)^{1/2}$ in the
weak-field limit to
\begin{equation}
l_{\text{jv}}(g_\ell,g_\delta)=\dfrac{\pi\sqrt{\left(g_{\delta}+
g_{\ell}\right)\lambda_{\text{L}}\xi}}{\ln^{1/2}\left[1+
\dfrac{4g_{\ell}\left(g_{\delta}+g_{\ell}\right)}{g_{\delta}^2}\right]},
\label{ljvPhi2}
\end{equation}
when half of the flux quantum pierces the junction.

In strongly anharmonic junctions ($g_{\ell}\gg g_\delta$) the
relation $l_{\text{jv}}\gg l_{\text{j}0}$ takes place, which
signifies a pronounced magnetic field dependence $l_{\text{j}}(\Phi)
$. The corresponding quantitative condition follows from
\eqref{ljvPhi2}:
\be
l_{\text{jv}}\approx \dfrac{g_{\ell}}{g_\delta}
\dfrac{\pi}{\sqrt{2}\ln^{1/2}\frac{2g_\ell}{g_\delta}}l_{\text{j}0}\gg
l_{\text{j}0}.
\label{ljvgglj0}
\ee

A significant difference between the characteristic size of the
Josephson vortex $l_{\text{j}}(\frac{\Phi_0}{2})=l_{\text{jv}}$ and
the weak-field penetration depth $l_{\text{j}0}$ is in striking
contrast with the harmonic junctions, where $l_{\text{jv}}=
\frac{\pi}{2}l_{\text{j}0}=\frac{\pi}{2}\lambda_{\text{J}}$.

A substantial increase of $l_{\text{jv}}$ as compared to
$l_{\text{j}0}$ is associated with the behavior of the quantity
$\int\nolimits_{0}^{\pi}j(\chi)d\chi$, which enters the right hand
side of \eqref{deflambdaj2} at $\Phi=\frac{
\Phi_0}{2}$. For the harmonic current $\int\nolimits_{0}^{\pi}j(\chi)
d\chi=2j_{\text{c}}$ that leads to $l_{\text{jv}}=\frac{\pi}{2}
l_{\text{j}0}$. In the strongly anharmonic regime, when $g_\ell\gg
g_\delta$ and $g_\delta^2\gg1$, the Josephson current $j(\chi)$
is small outside a pronounced narrow peak of the width $\Gamma\sim
\frac{g_\delta}{g_\ell}\ll 1$ in a vicinity of $\chi=\chi_{c}$ (see
(S11), (S12) in \cite{suppl} and also Fig.~2 in Ref.~\onlinecite{Barash2012}). The
critical current $j_{\text{c}}$ is determined by the height of the
peak: $j_{\text{c}}=j(\chi_c)$. Therefore, a qualitative estimate is
$\int\nolimits_{0}^{\pi} j(\chi)d\chi\sim \Gamma j_{\text{c}}\ll
j_{\text{c}}$. In accordance with Eq.~\eqref{deflambdaj2}, this
results in an increase of $l_{\text{jv}}$. Such an unconventional
behavior of $j(\chi)$ originates from the phase-dependent proximity
effect near the interface, which takes place when $g_\ell\gg
g_\delta$ and $g_\delta^2\gg1$~\cite{suppl}.

\begin{figure}[t]
\includegraphics[width=0.8\columnwidth,clip=true]{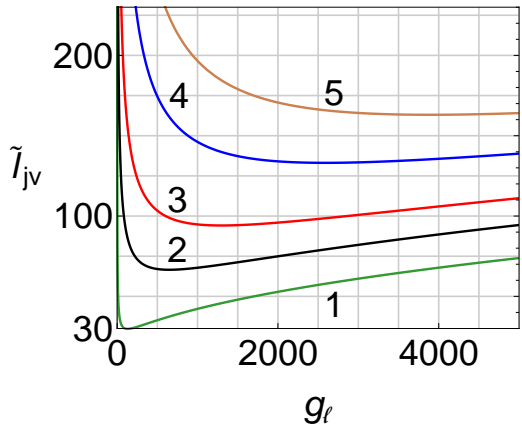}
\caption{The dimensionless characteristic size of the Josephson vortex
$\tilde{l}_{\text{jv}}$ as a function of $g_{\ell}$ taken for
various $g_\delta\gg1$:\,\, (1)\, $g_{\delta}= 100,$\,\,
(2)\, $g_{\delta}=500,$\,\, (3)\, $g_{\delta}=1000,$\,\,
(4)\, $g_{\delta}=2000,$\, and (5)\, $g_{\delta}=3000$.}
\label{lambdagell}
\end{figure}

To ensure the applicability of the result \eqref{ljvgglj0},
the conditions $g_\ell\gg g_\delta$,\, $g_\delta^2\gg1$, allowing
strongly anharmonic effects to manifest themselves in the junctions,
have to be restricted further as the consequence of applying the
local electrodynamics. This leads to the relation $l_{\text{j}}(\Phi)
\gg\lambda_{\text{L}}$, which is sensitive to the magnetic flux. In
weak fields one gets $l_{\text{j}0}\gg \lambda_{\text{L}}$ and
ultimately $\kappa^{1/2} g_\ell^{1/2}\ll g_\delta$. Joining the
conditions, results in strong inequalities $\kappa^{1/2}g_\ell^{1/2}
\ll g_\delta\ll g_{\ell}$,\, $g_\delta^2\gg1$. If these are
satisfied, the relation $l_{\text{j}}(\Phi)\gg\lambda_{\text{L}}$ and
the results \eqref{ljPhi2}-\eqref{ljvgglj0} would take place in the
whole region $|\Phi|\le\frac12\Phi_0$. The conditions $\kappa^{1/2}
g_\ell^{1/2}\ll g_\delta\ll g_{\ell}$ are quite restrictive and
uncommon as they can be satisfied only at huge values of $g_\ell$. On
account of a monotonic increase of $l_{\text{j}}(\Phi)$ with $\Phi$,
substantially weaker conditions $l_{\text{jv}}\gg\lambda_{\text{L}}$
emerge at $|\Phi|=\frac12\Phi_0$. Then the local approach is
justified in describing the Josephson vortex and, in particular, its
size \eqref{ljvPhi2}, and can fail at smaller values of $|\Phi|$. This
results in $g_\ell^{1/2}\gg\kappa^{1/2}$, up to a logarithmic factor.

\begin{figure}[t]
\includegraphics[width=0.815\columnwidth,clip=true]{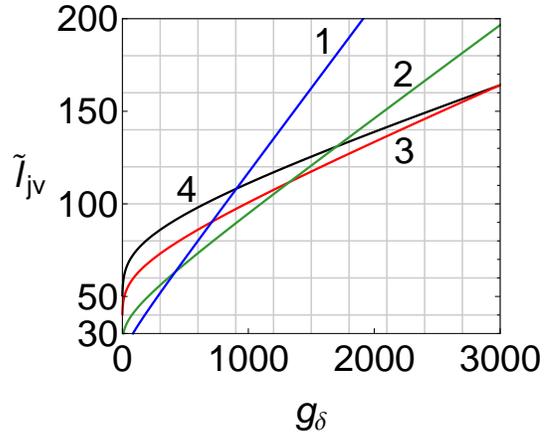}
\caption{The dimensionless characteristic size of the Josephson
vortex $\tilde{l}_{\text{jv}}$ as the function of $g_{\delta}$ taken
for various $g_\ell$:\,\, (1)\, $g_{\ell}= 300,$\,\,
(2)\, $g_{\ell}=1000,$\,\, (3)\, $g_{\ell}=3000,$\, and
(4)\, $g_{\ell}=5000$.}
\label{lambdagdelta}
\end{figure}

As seen from \eqref{ljvPhi2}, the dimensionless characteristic size
of the Josephson vortex $\tilde{l}_{\text{jv}}(g_{\ell},g_{\delta})=
l_{\text{jv}}\bigl/\sqrt{\lambda_{\text{L}}\xi}$ is expressed, within
the GL theory, solely via the parameters $g_\ell$ and $g_\delta$. The
quantity $\tilde{l}_{\text{jv}}$ is shown in Fig.~\ref{lambdagell}
for various strengths of the interfacial pair breaking, as a function
of $g_\ell$. The numerical results have been obtained by
carrying out the evaluation of $j(\chi)$ with the exact
self-consistent formulas of the GL theory \cite{Barash2012,suppl},
and further by calculating the junction penetration depth
\eqref{deflambdaj2} at $\Phi=\frac12\Phi_0$. All the curves in
Fig.~\ref{lambdagell} are perfectly approximated by
Eq.~\eqref{ljvPhi2}. As a function of the Josephson coupling strength
$g_\ell$, the quantity $l_{\text{jv}}$ shows a nonmonotonic behavior.
In the harmonic regime $g_\ell\ll g_\delta$ the penetration depth
decreases with $g_\ell$ as $l_{\text{j}}\propto g_{\ell}^{-1/2}$ (see
\eqref{lambdaj1}). When the parameter $g_\ell$ increases further and
the anharmonic features of the current-phase relation become
pronounced, the integral of the supercurrent over the phase
difference in Eq.~\eqref{deflambdaj2} diminishes, as discussed above.
As a consequence, the junction penetration depth \eqref{deflambdaj2}
(and, in particular, \eqref{ljPhi2}, \eqref{ljvPhi2}) gradually goes
up with increasing $g_{\ell}$ in the region $g_{\ell}\gg g_\delta$,
$g_{\delta}^2\gg1$. As follows from \eqref{ljvPhi2}, a minimum of
$l_{\text{jv}}(g_\ell, g_\delta)$ as a function of $g_\ell$ at fixed
$g_\delta$ takes place at $g_{\ell}^2\sim g_{\delta}^2\gg 1$.
Specifically, the minima of $\tilde{l}_{\text{jv}}(g_{\ell},
g_{\delta})$, which correspond to the curves 1 and 2 of
Fig.~\ref{lambdagell}, are $l_{\text{jv},\text{min}} (g_\delta=100)
\approx 29.7675$ at $g_\ell\approx 129.562$, and $l_{\text{jv},
\text{ min}}(g_\delta=500) \approx 66.5611$ at $g_\ell\approx
647.781$.

After rewriting the condition $l_{\text{jv}}(g_{\ell},g_{\delta})
\gg\lambda_{\text{L}}$ in the form $\tilde{l}_{\text{jv}}(g_{\ell},
g_{\delta})\gg \sqrt{\kappa}$, one can see that the applicability
domain of the results shown in the figures and obtained within the
local theory, depends on the GL parameter $\kappa\gg1$. For example,
all the curves in Fig.~\ref{lambdagell} satisfy the condition
$\tilde{l}_{\text{jv}}(g_{\ell},g_{\delta})\ge 30\gg 3$ and,
therefore, they are applicable to the case $\kappa=10$. However, for
$\kappa=100$ a substantial part of curve 1 does not satisfy the
condition $\tilde{l}_{\text{jv}}(g_{\ell},g_{\delta})\gg 10$ in the
given region of $g_\ell$. It is in contrast to other curves,
which remain wholly justified. Similar remarks would apply in fact
to all subsequent figures of the paper (Figs.~\ref{lambdagdelta}-\ref{hc1h0}).

With increasing the interface parameter $g_{\delta}$, the critical
current $j_{\text{c}}$ and the integral $\int\nolimits_{0}^{\pi}
j(\chi)d\chi$ decrease irrespective of the relation between
$g_{\delta}$ and $g_{\ell}$. For this reason and in accordance with
\eqref{deflambdaj2} and \eqref{ljPhi2}, the junction penetration
depth monotonically increases with increasing $g_{\delta}$ at fixed
$\Phi$ and $g_\ell$. The quantity $\tilde{l}_{\text{jv}}$ as a
function of $g_{\delta}$, taken for various $g_\ell$, is depicted in
Fig.~\ref{lambdagdelta}. In the region $g_\delta^2\gg 1$ the exact
results are in agreement with those following from \eqref{ljvPhi2}.
As a consequence of the nonmonotonic dependence on $g_\ell$, the
curves in Fig.~\ref{lambdagdelta}, which correspond to different
$g_\ell$, can cross each other.

\section{The spatial structure of an isolated Josephson vortex}
\label{sec: josscreen}

\begin{figure}[t]%[thb!]
\includegraphics[width=0.715\columnwidth,clip=true]{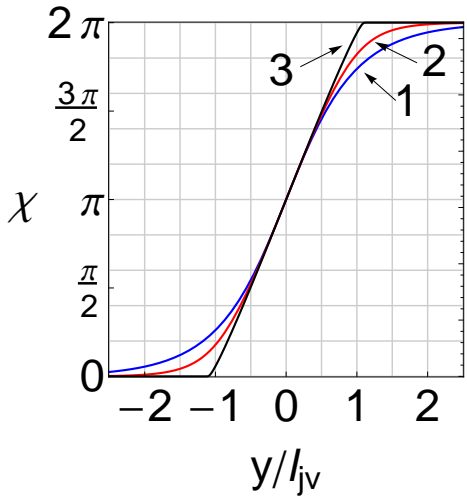}
\caption{The spatial profile of the phase difference in the
Josephson vortex in the junctions with $g_\delta=10^2$ and
various $g_\ell$:\,\, (1)\, small Josephson couplings $g_{\ell}\ll 1,$
\,\, (2)\, $g_{\ell}=10^2,$\, and (3)\, $g_{\ell}=10^4$.}
\label{phase}
\end{figure}

While in harmonic junctions the Josephson screening of
the magnetic field is characterized by the only length scale
$\lambda_{\text{J}}$, in the strongly anharmonic junctions the
spatial distributions of $\chi(y)$, $H(y)$ and $j(y)$ along the
junction plane contain two characteristic lengths, at a fixed value
of the applied magnetic field. Since in the strongly anharmonic
regime the current density $j(\chi)$ has a pronounced narrow peak as
a function of $\chi$, the Josephson current experiences abrupt
spatial changes in a small region of $y$, where varying in space
phase difference passes through the vicinity of $\chi(y_c)= \chi_c$
with a change of $y$. The quantities $H(y)$ and
$\frac{d\chi(y)}{dy}$ change comparatively quickly in that small
space region, so that a smaller characteristic length is determined
by the particular form of the anharmonic current-phase relation. As
a result, the spatial profile of $j[\chi(y)]$ contains narrow peaks,
while $\chi(y)$ and $H(y)$ acquire more angular shape as compared to
that in the harmonic junctions. The greater characteristic length is
the junction penetration depth $l_{\text{j}}(\Phi)$, which can evolve
considerably with the varying magnetic field, as shown in the
preceding section. Due to a small value of the supercurrent outside
the peak region, in strongly anharmonic junctions
$\frac{d\chi(y)}{dy}$ is almost constant and $\chi(y)$ is nearly a
linear function of $y$ over the scale $\sim l_{\text{j}}(\Phi)$.

\begin{figure}[t]%[thb!]
\includegraphics[width=0.815\columnwidth,clip=true]{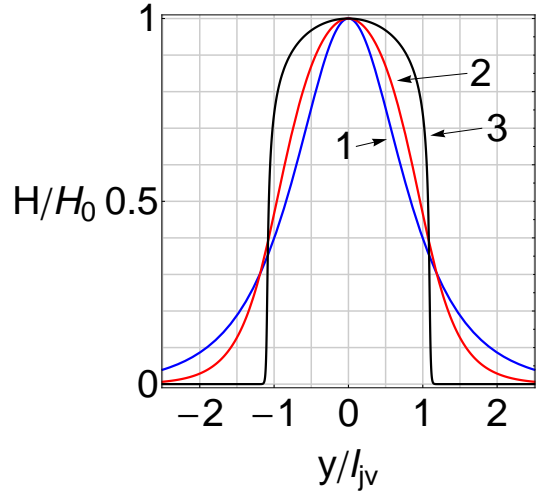}
\caption{The spatial profile $H(\tilde{y})$ in the Josephson
vortex, normalized to the field value $H_0$ in its center,
in junctions with $g_\delta=10^2$ and various $g_\ell$:\,\,
(1)\, small Josephson couplings $g_{\ell}\ll 1,$\,\, (2)\,
$g_{\ell}=10^2,$\, and (3)\, $g_{\ell}=10^4$.}
\label{field}
\end{figure}

Consider here an isolated Josephson vortex deep inside the junction
plane, which is known to contain a single flux quantum. The magnetic
field is symmetric and the current density is antisymmetric with
respect to the vortex center, while $\chi(y)$ changes monotonically
overall by $2\pi$. In the presence of a strong interfacial pair
breaking $g_\delta^2\gg 1$, the equation describing the spatial
dependence of the phase difference within the local Josephson
electrodynamics can be written as \cite{suppl}
\begin{equation}
\dfrac{d\chi}{d\tilde{y}}=\pi\,\dfrac{\ln^{1/2}\left[1+
\dfrac{4g_{\ell}\left(g_{\delta}+g_{\ell}\right)}{g_{\delta}^2}
\sin^2\dfrac{\chi(\tilde{y})}{2}\right]}{\ln^{1/2}\left[1+
\dfrac{4g_{\ell}\left(g_{\delta}+g_{\ell}\right)}{g_{\delta}^2}\right]}.
\label{derchi2}
\end{equation}
Here the dimensionless coordinate $\tilde{y}={y}\big/{l_{\text{jv}}}$ is
introduced and $H(y)>0$ assumed.

\begin{figure}[t]
\includegraphics[width=0.815\columnwidth,clip=true]{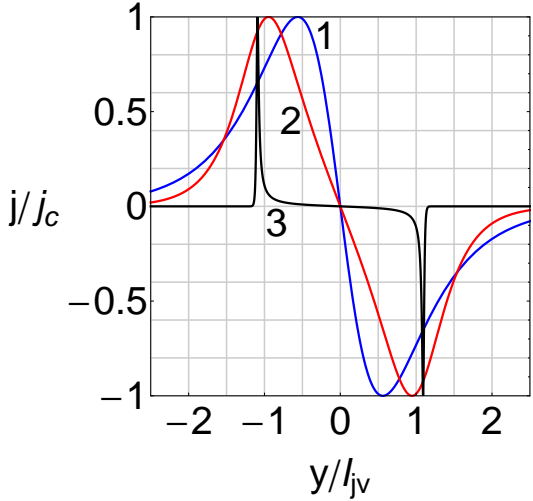}
\caption{The spatial profile $j(\tilde{y})$, normalized to its
critical value $j_{\text{c}}$, in the Josephson vortex in junctions
with $g_\delta=10^2$ and various $g_\ell$:\,\, (1)\, small Josephson
couplings $g_{\ell}\ll 1,$\,\, (2)\, $g_{\ell}=10^2,$\, and
(3)\, $g_{\ell}=10^4$.}
\label{current}
\end{figure}

The spatial profiles of the phase difference $\chi(\tilde{y})$, of
the magnetic field $H(\tilde{y})$ and of the supercurrent density
$j(\tilde{y})$ in an isolated Josephson vortex are depicted in
Figs.~\ref{phase}-\ref{current}. The vortex center is taken here at
$y=0$, and the asymptotic values of the phase difference are
$\chi_{-\infty}=0$,\, $\chi_{\infty}=2\pi$. All the distributions
obtained confirm that the overall large scale of the spatial
variations is $l_{\text{jv}}$. In this respect, even a comparatively small
difference between $\lambda_{\text{J}}$ and $l_{\text{jv}}=\frac{\pi}{2}
\lambda_{\text{J}}$ in harmonic junctions can be discerned. In each
of these figures curve 1 describes the behavior of the
corresponding quantity in harmonic junctions with a strong
interfacial pair breaking. It is similar to the analogous profiles in
standard tunnel junctions. A particularly interesting case of
strongly anharmonic junctions is described by curve 3 in the
figures. Curve 3 in Fig.~\ref{phase} demonstrates a
sharp crossover of the gradual behavior of the phase difference and
its asymptotic value. It is in contrast to curve 1, which
smoothly varies over the only scale $l_{\text{jv}}$. Curve 2
shows an intermediate behavior. Similarly, in contrast to curves
1 and 2, curve 3 in Fig.~\ref{field}, which describes the profile
of the magnetic field in strongly anharmonic junctions, shows no
noticeable tails of the field at distances $|y|>l_{\text{jv}}$.

Curve 3 in Fig.~\ref{current} demonstrates, that the supercurrent
flows in strongly anharmonic junctions mostly in a small narrow part
of the Josephson vortex. As was noted above, these are the narrow
peaks in the current-phase relation of strongly anharmonic junctions,
which transform into spatial peaks of the supercurrent density
due to a spatial dependence of the phase difference. An effect of
similar origin, but with a transformation into the magnetic flux
dependence, has been recently predicted in strongly anharmonic
junctions, whose widths are much less than the junction penetration
depth.~\cite{Barash2012_3} The narrow central Fraunhofer peak of the
total critical current was found to possess the following half width
at the half of the peak $(\Delta\Phi\bigl/\Phi_0)\approx 1.35
g_\delta\bigl/g_\ell\ll 1$, under the conditions $g_\ell\gg g_\delta$
and $g_\delta^2\gg1$.

\section{The lower critical field}
\label{sec: lowerfield}

The lower critical field of the junction is known to satisfy the
relation $H_{\text{jc}1}=4\pi {\mathit\varOmega}_{l}\bigl/\Phi_0$,
where ${\mathit\varOmega}_{l}$ is the thermodynamic potential of
the Josephson vortex per unit length.
For the junctions with an intense interfacial pair
breaking $g_\delta^2\gg 1$ one gets \cite{suppl}
\be
H_{\text{jc}1}=\dfrac{\Phi_0g_\delta\int\limits_{-\infty}^{\infty}
\ln\left[1+\frac{4g_\ell(g_\delta+g_\ell)}{g_\delta^2}
\sin^2\frac{\chi(\tilde{y})}{2}\right]d\tilde{y}}{8\pi\lambda_L
\lambda_{\text{J}}\sqrt{g_\ell(g_\delta+g_\ell)
\ln\left[1+\frac{4g_\ell(g_\delta+g_\ell)}{g_\delta^2}\right]}},
\label{hc12}
\ee
where $\chi(\tilde{y})$ is the solution of Eq.~\eqref{derchi2} and
$\lambda_{\text{J}}\approx (\lambda_L\xi\bigl/g_\ell)^{1/2}g_\delta$
(see \eqref{lambdaj1}).
For harmonic junctions, when $g_\ell\ll g_\delta$, the
logarithmic functions in \eqref{hc12} can be expanded
and the integral calculated with the solution $\chi(\tilde{y})=-2
\arcsin\sech\bigl(\pi \tilde{y} \bigl/2\bigr)$. This results in
$H_{\text{jc}1}=\Phi_0\bigl/(\pi^2 \lambda_L\lambda_{\text{J}})$,
in agreement with the conventional expression.~\cite{Josephson1965,%
Tinkham1996}

\begin{figure}[t]
\includegraphics[width=0.77\columnwidth,clip=true]{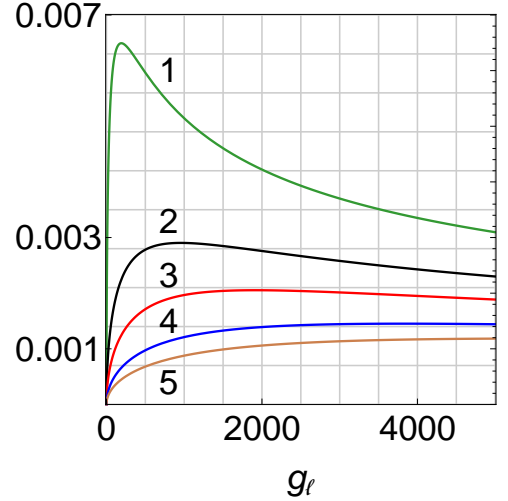}
\caption{The dimensionless lower critical field
$\tilde{H}_{\text{jc}1}$ as a function of $g_{\ell}$, taken for
various $g_\delta\gg1$:\,\, (1)\, $g_{\delta}= 100,$\,\,
(2)\, $g_{\delta}=500,$\,\, (3)\, $g_{\delta}=1000,$\,\,
(4)\, $g_{\delta}=2000,$\, and (5)\, $g_{\delta}=3000$.}
\label{hc1h*}
\end{figure}

To single out the dependence of $H_{\text{jc}1}$ on the effective
interface parameters $g_\ell$ and $g_\delta$, it is convenient to
introduce the dimensionless lower critical field of the junctions
$\tilde{H}_{\text{jc}1}=H_{\text{jc}1}\bigl/H^*$, taken in units of
$H^*=\Phi_0\Bigl/(\lambda_L^{3/2}\xi^{1/2})$. Figure~\ref{hc1h*}
displays $\tilde{H}_{\text{jc}1}$ as a function of the strength of
the Josephson coupling, for various values of the strong interfacial
pair breaking. It is a nonmonotonic function of $g_\ell$. In tunnel
junctions the field increases with $g_\ell$ and decreases with
$g_\delta$ as $H_{\text{jc}1}=2\Phi_0\bigl/(\pi^2\lambda_J
2\lambda_L)\propto g_\ell^{1/2}g_\delta^{-1}$. A decrease with
$g_\ell$ under the conditions $g_\ell\gg g_\delta\gg 1$ takes place
for the same reason, for which the penetration depth increases (see
Fig.~\ref{lambdagell} above and its discussion).

\begin{figure}[t]
\includegraphics[width=0.82\columnwidth,clip=true]{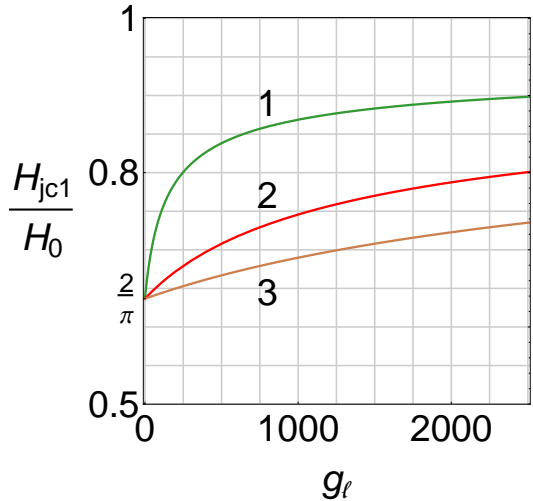}
\caption{The ratio $H_{\text{jc}1}\bigl/H_0$ as a function of $g_{\ell}$,
for various $g_\delta\gg1$:\,\, (1)\, $g_{\delta}= 100,$\,\, 
(2)\, $g_{\delta}=1000,$\, and (3)\, $g_{\delta}=3000$.}
\label{hc1h0}
\end{figure}

The relation $H_{\text{jc}1}\ll H_{\text{c}1}
$ is to a large extent close to the strong inequality $l_{\text{jv}}
\gg\lambda_L$, which determines the applicability of the local theory
to describing the structure of the Josephson vortex. Here
$H_{\text{c1}}=\Phi_0\bigl(\ln\kappa+0.08\bigr)\Bigl/\bigl(4\pi
\lambda_L^2\bigr)$ is the lower critical field of the massive
strongly type-II superconductor without weak links. After reducing the
relation $H_{\text{jc}1}\ll H_{\text{c}1}$ to the form
$\tilde{H}_{\text{jc}1}\ll(\ln\kappa+0.08)\bigl/(4\pi\sqrt{\kappa})$,
one sees that the applicability domain of the condition
$H_{\text{jc}1}\ll H_{\text{c}1}$, in terms of Fig.~\ref{hc1h*},
depends on the GL parameter $\kappa\gg1$. For example, for $\kappa=10
$ one obtains a strong inequality $\tilde{H}_{\text{jc}1}\ll 0.06$,
which applies to all the curves in Fig.~\ref{hc1h*}. For $\kappa=100$
one gets the relation $\tilde{H}_{\text{jc}1}\ll 0.037$. It is fully
applicable to curves 2-5 and only partially to curve 1. This is
very similar to what was said above regarding Fig.~\ref{lambdagell}
plotted for the dimensionless quantity $\tilde{l}_{\text{jv}}$, taken
for the same set of parameters.

In tunnel junctions, the field $H_0$ in the center of the
Josephson vortex is related to $H_{\text{jc}1}$ as
$H_{\text{jc}1}= \frac{2}{\pi}H_0$.~\cite{Josephson1965,Tinkham1996}
In strongly anharmonic junctions with an intense interfacial pair
breaking, the ratio $H_{\text{jc}1}\bigl/H_0$ depends on $g_\ell$ and
$g_\delta$. As shown in Fig. \ref{hc1h0}, the ratio varies between
$\frac{2}{\pi}$ and unity.  It monotonically increases with $g_\ell$,
while the pair breaking tends to suppress it towards its standard
value.

\section{Discussion}
\label{sec: discussion}

A pronounced unconventional behavior of the magnetic properties of
the planar junctions emerges, when the parameters $g_\ell$ and
$g_\delta$ are large and satisfy the conditions $g_\delta^2\gg1$,
$g_\ell\gg g_\delta$. A phase dependent suppression of the order
parameter at the junction interface, taking place under such
conditions due to the proximity effect (see (S13) in \cite{suppl}), is of key
importance here. Qualitatively, the Josephson current (S11)
increases and the junction penetration depth \eqref{ljPhi2},
\eqref{ljvPhi2} decreases with increasing $g_\ell$, when $g_\ell
\ll g_\delta$, $g_\delta^2\gg1$ and the suppression does not
substantially depend on $g_\ell$. However, if $g_\ell\gg g_\delta$,
then a phase dependent local decrease of the condensate density at
the interface takes place with $g_\ell$ and results in the
anharmonic Josephson current, which substantially decreases in the
wide region of the phase difference. Though the critical current does
not diminish in this regime (see (S12) in \cite{suppl}), the integral
$\int\nolimits_{0}^{\pi}j(\chi)d\chi$ decreases with $g_\ell$. This
induces an increase of the Josephson vortex size $l_{{jv}}$,
demonstrating an important role the anharmonic effects play in the
problem in question.

A possibility of achieving large values of $g_\ell$ and $g_\delta$
in experiments has not been established as yet. However, a number of
microscopic models persuasively indicate that the strong inequalities
$g_\delta^2\gg1$, $g_\ell\gg g_\delta$, resulting in a pronounced
anharmonic current-phase relation, can be satisfied under certain
conditions. More restrictive relations emerge due to the application
of local Josephson electrodynamics at large $g_\ell$. In weak fields
this results in the uncommon conditions $\kappa^{1/2}g_\ell^{1/2}\ll
g_\delta\ll g_{\ell}$, which are only satisfied at huge values $g_\ell
\agt 10^4$, and could be challenging in an experimental realization.
For the Josephson vortices the conditions are substantially weaker:\,
$g_\ell^{1/2}\gg\kappa^{1/2}$,\, $g_\delta^2\gg1$,\, $g_\ell\gg g_\delta$.

There are no fundamental upper bounds to large values of the
parameters $g_\ell$ and $g_\delta$. Microscopic model results for
$g_\ell$ can be obtained based on the corresponding studies of the
Josephson current near $T_c$ \cite{Ivanov1981,Kupriyanov1992,%
Golubov2004,Barash2012_2}, or of the boundary conditions for the
superconductor order parameter at the interface in the GL theory \cite{deGennes1966,%
Galaiko1969,Bratus1977,Svidzinskii1982,Geshkenbein1988}. As follows
from the microscopic results, in dirty junctions with small and
moderate transparencies, $g_\ell$ can vary from vanishingly small
values in the tunneling limit to those well exceeding $10^2$ and
leading to a substantially anharmonic behavior of the Josephson
current. The parameter $g_\ell$ goes up, when the interface
transparency increases. In highly transparent planar junctions $g_\ell$ can
generally take huge values. As $g_\ell\propto\xi(T)$, an additional
increase of $g_\ell$ occurs near $T_c$.

Large $g_\delta$ corresponds to a strong suppression of
the order parameter at the junction interface.
A strong interfacial pair breaking can be induced by
proximity to superconductor-normal metal interfaces and to
magnetically active boundaries in various superconductors, including
isotropic $s$-wave ones.~\cite{deGennes1964,ROZaitsev1965,%
deGennes1966,Ivanov1981,Sauls1988,Nazarov2009} In unconventional
superconductors a significant pair breaking can be present also near
superconductor-insulator and superconductor-vacuum
interfaces.~\cite{Buchholtz1981,Barash1995,Shiba1995a,*Shiba1995b,%
*Shiba1995c,*Shiba1996,Nagai1995,Sauls1995a,*Sauls1995b,Alber1996,%
Agterberg1997} Under certain conditions, the order parameter
can be fully suppressed on the boundary, in particular, for symmetry
reasons in unconventional superconductors. This signifies that
$g_\delta$ can, in general, take huge values.

A specific microscopic example studied in detail theoretically
\cite{Ivanov1981,Golubov2004}, in which the parameters $g_\ell$ and
$g_\delta$ of the GL theory can satisfy the conditions $g_\delta^2
\gg1$, $g_\ell\gg g_\delta$, is the dirty SNS junction where the
normal conductivity of the leads is significantly less than the
conductivity of a thin normal metal interlayer.

\section{Conclusions}
\label{sec: conclusions}

The problem of the magnetic field screening and of the Josephson
vortex structure in superconducting planar junctions with anharmonic
current-phase relations, has been solved in this paper within the GL
theory. Since a strongly anharmonic behavior only appears due to a
pronounced Josephson coupling, an intense interfacial pair breaking
needs to be present for the planar junctions to be weak links.
Another reason for a pronounced interfacial pair breaking to play a
crucial role for the theory developed, is that an intense pair
breaking significantly increases the penetration depth and thereby
substantially extends an applicability domain of the local
Josephson electrodynamics, which in this case applies to the
junctions with the strong Josephson coupling.

The magnetic penetration depth $l_{\text{j}}$ in the junctions
is identified theoretically as a function of the magnetic
flux, of the Josephson coupling strength and the interfacial pair
breaking. Due to a nonexponential spatial profile of the screened
magnetic field in the junction plane, a quantitative definition of
$l_{\text{j}}$ is put to use, similar to the standard definitions of
magnetic penetration depths in various other circumstances. In
harmonic junctions a characteristic size of the Josephson vortex
along the junction plane $l_{\text{jv}}$ and the weak-field
penetration depth $l_{\text{j0}}$ are shown to be related as
$l_{\text{jv}}=\frac{\pi}{2}l_{\text{j0}}$. A pronounced magnetic
field dependence of $l_{\text{j}}$, which induces a significant
increase of $l_{\text{jv}}$ as compared to $l_{j0}$, is predicted
in strongly anharmonic junctions. A nonmonotonic dependence of
$l_{\text{jv}}$ on the Josephson coupling strength is obtained in
such junctions, as a consequence of the phase-dependent proximity
effect, and demonstrated to result in an applicability of
the local approach to sufficiently large $g_{\ell}$ at a fixed
$g_{\delta}$.

A narrow peak in an anharmonic current-phase relation
was found to induce a peak in the spatial profile of the
supercurrent density, which is narrow compared to the size of the
Josephson vortex. A nonmonotonic dependence on the Josephson coupling
as well as a monotonic one on a strength of the interfacial
pair breaking, was obtained for the lower critical field of the
junctions.

An inclusion of the nonlocal effects into the theory developed above
is desirable. Possible restrictions on the results obtained could be
also associated with imperfections of the planar geometry of the
junctions. Their study would require a significant extension of the
theoretical approach used, and lies outside the scope of the paper.

\begin{acknowledgments}
The support from Russian Foundation for Basic Research under grants 11-02-00398
and 14-02-00206 is acknowledged.
\end{acknowledgments}

\providecommand{\noopsort}[1]{}\providecommand{\singleletter}[1]{#1}%
%

%%%%%%%%%%%%%%%%%%%%%%%%%%%%%%%%%%%%%%%%%%%%%
%% Supplemental material
%%%%%%%%%%%%%%%%%%%%%%%%%%%%%%%%%%%%%%%%%%%%%

\pagebreak
\onecolumngrid
\vspace{0.2in}
\begin{center}
{\bf \large  Magnetic penetration depth and vortex structure in anharmonic superconducting junctions with an interfacial 
pair breaking - Supplemental material}
\end{center}
\vspace{0.1in}

\renewcommand{\thesubsection}{S\arabic{subsection}}
\renewcommand{\theequation}{S\arabic{equation}}

\setcounter{secnumdepth}{2}
\setcounter{equation}{0}
\setcounter{figure}{0}
\setcounter{section}{0}

In this supplemental material I present self-consistent results
of the GL theory for the Josephson current in planar junctions as a
function of the Josephson coupling, of the interfacial pair breaking
and of the phase difference. Also the derivations of the defining
expressions are given for the magnetic penetration depth and for the
lower critical field in anharmonic junctions. The first integral of
the basic differential equation for the spatially dependent phase
difference is derived for the Josephson vortex in anharmonic
junctions.

\vspace{0.5cm}

\twocolumngrid

\subsection{The Josephson current as a function of $g_\ell$,~$g_\delta$~and~$\chi$}

The GL free energy of the junction is represented as a sum of three
terms:\, ${\cal F}={\cal F}_{b1}+{\cal F}_{b2}+{\cal F}_{\text{int}}
$.  For symmetric junctions between two identical $s$-wave or
$d_{x^2-y^2}$-wave superconductors, the bulk and the interface free
energies can be written, in the absence of a magnetic field, as
\begin{multline}
{\cal F}_{b1(2)}=\!\!\int\nolimits_{V_{1(2)}}\!\Bigl(
K\left|\pmb\nabla\Psi_{1(2)}\right|^2+
a\left|\Psi_{1(2)}\right|^2+\\ +
\dfrac{b}{2}\left|\Psi_{1(2)}\right|^4\Bigr)dV_{1(2)},
\label{fb1}
\end{multline}
\begin{eqnarray}
{\cal F}_{\text{int}}=\!\!\!\int\nolimits_{S}\Bigl[
g_{J}\left|\Psi_1-\Psi_2\right|^2\!\!
+g\left(\left|\Psi_{1}\right|^2\!\!+\left|\Psi_2\right|^2\right)\Bigr]dS\,.
\label{fint1}
\end{eqnarray}

The interface free energy \eqref{fint1} contains both the Josephson
coupling of the superconducting banks with the coupling constant
$g_J$, and the term with the coupling constant $g$, which describes,
for instance in the absence of the current, the interfacial pair
breaking ($g>0$), or the pair formation ($g<0$).

The two independent interface invariants in \eqref{fint1}, which
control jointly the relative value of the supercurrent with respect
to the depairing current $j_{\text{dp}}$, are qualitatively
different. In the absence of the current, i.e., at $\chi=0$, one
gets $\Psi_{1}=\Psi_{2}$ in symmetric junctions and, therefore, the
first invariant in \eqref{fint1} vanishes. On the contrary, the
invariant $\Psi_{1}\Psi_{2}^*+\Psi_{1}^*\Psi_{2}$ does not vanish
at $\chi=0$. Therefore, if it were used for describing the Josephson
coupling instead of the first term in \eqref{fint1}, it would result
at $\chi=0$ in the same interfacial proximity effect as the second
invariant in \eqref{fint1}. Such a duplication is microscopically
unjustified and would complicate the analysis of the problem. In
particular, the free energy \eqref{fint1}, in line with the
microscopic theory, satisfies the condition that there should be no
influence of a thin interface on the superconductors in symmetric
junctions with the vanishing interfacial pair activity ($g=0$) and
at the zero phase difference ($\Psi_1=\Psi_2$). The last statement is
related, to a certain degree, to Anderson theorem regarding a
negligible influence of nonmagnetic impurities on the thermodynamic
superconductor properties, in contrast to their profound effect on
the supercurrent flow and correlations.

Taking the order parameter in the form $\Psi_{1(2)}=(|a|/b)^{1/2}
f_{1(2)}(x)e^{i\chi_{1(2)}(x)}$, one can transform the GL equations,
which follow from the bulk free energies \eqref{fb1}, to the
equations for the normalized order-parameter moduli
\be
\dfrac{d^2f_{1(2)}}{d\overline{x}^2}-\dfrac{4\tilde{\jmath}^2}{27f^3_{1(2)}}+f_{1(2)}-f_{1(2)}^3=0
\label{gleq2}
\ee
and to the current conservation condition. Here $\overline{x}=x/\xi(T)$,
$\xi(T)=(K/|a|)^{1/2}$ is the temperature dependent superconductor
coherence length, $\tilde{\jmath}={j}/j_{\text{dp}}=-(3\sqrt{3}/{2})
({d\chi}/{d\overline{x}})f^2$ is the spatially constant normalized
current density, $j_{\text{dp}}=\bigl(8|e||a|^{3/2}K^{1/2}\bigr)\big/
\bigl(3\sqrt{3}\hbar b\bigr)$ is the depairing current deep inside
the superconducting leads. As $df/dx=0$ in the bulk $x\to\pm\infty$,
one finds from \eqref{gleq2} the relation $\tilde{\jmath}^2=(27/4)
(1-f_{\pm\infty}^2)f_{\pm\infty}^4$. Here $\frac23<f_{\pm\infty}^2\le
1$ \cite{Tinkham1996} and, in particular, in the absence of the
supercurrent $f_{\pm\infty}=1$. For symmetric junctions with $f$
continuous through the interface $f=f(|x|)$, i.e., $f_2(x)=f_1(-x)$.

The interface terms in \eqref{fint1} and the gradient term in
\eqref{fb1} contribute to the boundary
conditions for complex order parameters. One can split them into the
boundary conditions for $f$ and the expression for the Josephson
current via the value $f_{0}$ at $x=0$ and the phase difference
$\chi=\chi_--\chi_+$ across the interface:
\begin{align}
\biggl(&\!\frac{df}{d\overline{x}}\!\biggr)_{\pm} =
\pm\!\left(g_{\delta}+2g_\ell\sin^2\!\frac{\chi}2\right)f_0,
\label{bcss99s}\\
&\tilde{j}=\frac{3\sqrt{3}}{2}g_\ell f_{0}^2\sin\chi.
\label{bcss9s}
\end{align}
Here $g_\ell=g_J\xi(T)/K$ is the effective dimensionless Josephson
coupling constant and $g_{\delta}=g\xi(T)/K$ is the effective
dimensionless interface parameter.

For identifying the Josephson current based on \eqref{bcss9s}, one should
know the self-consistent interface value $f_0$ as a function of the
phase difference $\chi$. The simplest way to obtain the exact result
of the GL theory for $f_0$ and, hence, for the Josephson current
through the junctions in question, is to make use of the first
integral of Eq.~\eqref{gleq2}, which can be written as
\begin{equation}
\left(\frac{df}{d\overline{x}}\right)^2+f^2\!-
\frac{1}{2}f^4\!+\,\dfrac{4\tilde{\jmath}^2}{27f^2}=
2f_{\infty}^2-\dfrac{3}{2}f_{\infty}^4.
\label{gl1d8}
\end{equation}

The quantity $f_0$ can be found without resorting to a spatially
dependent solution of \eqref{gl1d8} or \eqref{gleq2}. One puts $x=0$
in \eqref{gl1d8}, substitutes \eqref{bcss9s} for the current and
considers the resulting equation as a polynomial one with respect to
three unknown quantities $\left(\frac{df}{d\overline{x}}\right)_0^2$,
$f_0^2$ and $f_{\infty}^2$, for a given phase difference $\chi$. An
additional relationship between $f_0^2$ and $f_{\infty}^2$ is
obtained by equating the current \eqref{bcss9s} to its expression via
$f_{\infty}^2$ given above. Together with the boundary conditions
\eqref{bcss99s}, one gets three equations for three unknown
quantities, which are reduced to a single fourth-order polynomial
equation
\begin{equation}
2g_b^{2}(\chi)\alpha-
(1-\alpha)^2[1-\alpha(\alpha+2)g_\ell^{2}\sin^2\chi]=0
\label{alphaeq}
\end{equation}
for the quantity $\alpha$. It relates the values of the order
parameter taken at the interface and in the bulk to each other:
$f_0^2=\alpha f_{\infty}^2$. The quantity $g_b(\chi)$ in
\eqref{alphaeq} is defined as $g_b(\chi)=\left(g_{\delta}+2g_\ell
\sin^2\frac{\chi}2\right)$.

Equating the current \eqref{bcss9s} to that in the bulk results in
$f_{\infty}^2=1-g_\ell^2\sin^2\chi\alpha^2$ and therefore in $f_0^2
=\alpha\left(1-g_\ell^2\sin^2\chi\alpha^2\right)$. Substituting the
latter formula in \eqref{bcss9s}, one expresses the Josephson current
via $\alpha$:
\begin{equation}
j=\dfrac{3\sqrt{3}}{2}\alpha g_{\ell}\sin\chi(1-
\alpha^2 g_\ell^2\sin^2\chi)\,j_{\text{dp}}.
\label{jgl3}
\end{equation}

Simple numerical solution of \eqref{alphaeq} allows one to describe,
based on \eqref{jgl3}, the Josephson current as a function of $\chi$
and of the parameters $g_{\delta}$ and $g_{\ell}$.~\cite{Barash2012}
Two of the four solutions of \eqref{alphaeq} take complex values and,
therefore, are of the unphysical character.
One of the two remaining solutions satisfies the condition $\alpha<1$
and corresponds to the pair breaking effects in the presence of the
current ($g_b(\chi)>0$). And finally, the fourth solution exceeds the
unity $\alpha>1$ and can be related to the junctions with the pair
forming interfaces ($g_b(\chi)<0$). Further on only the junctions
with $g_\delta, g_\ell>0$ will be considered.

In addition to the exact numerical results, one can also obtain an
analytical solution, which describes the Josephson current with
a good accuracy. With the pair breaking solution, the following
anharmonic current-phase relation has been obtained for the Josephson
current in the absence of the magnetic field~\cite{Barash2012}
\begin{multline}
j\left(g_{\ell},g_{\delta},\chi\right)=
\frac{3\sqrt{3}g_{\ell}\sin\chi}{2(1+2g_{\ell}^{2}
\sin^2\chi)}\biggl[1+g_b^{2}(\chi)+
g_{\ell}^{2}\sin^2\chi-\\
-\sqrt{\bigl(g_b^{2}(\chi)+g_{\ell}^{2}\sin^2\chi\bigr)^2+
2g_b^{2}(\chi)}\,\biggr]j_{\text{dp}}.
\label{bcss1012p}
\end{multline}

The exact numerical solution shows that
Eq.~\eqref{bcss1012p} describes the current behavior almost
perfectly, if $j< 0.7j_{\text{dp}}$. This concerns, in particular,
the current at $g_{\ell}<1$ for any $g_{\delta}$, or at $g_{\delta}>1
$ for any $g_{\ell}$. For $j>0.7j_{\text{dp}}$ Eq.~\eqref{bcss1012p}
represents a good approximation of the exact numerical solution, with
the deviations not exceeding $10\%$.

There are two basic limiting cases, when the expression
\eqref{bcss1012p} for the supercurrent considerably simplifies.
For tunnel junctions $g_\ell\ll 1$ in the
presence of an interfacial pair-breaking, the harmonic current-phase
relation is
\begin{equation}
j\left(g_{\ell},g_{\delta},\chi\right)=\dfrac{3\sqrt{3}}{4}g_\ell
\Bigl(\sqrt{g_\delta^2+2}-g_\delta\Bigr)^2j_{\text{dp}}\sin\chi\, .
\label{jtun1}
\end{equation}

On the other hand, in the regime of a pronounced interfacial pair
breaking $g_{\delta}^2\gg 1$ the anharmonic supercurrent at arbitrary
values of $g_{\ell}$ takes the form
\begin{equation}
j\left(g_{\ell},g_{\delta},\chi\right)=\frac{3\sqrt{3}g_\ell
j_{\text{dp}}\sin\chi}{4[g_\delta^2+4(g_\delta+g_\ell)g_\ell
\sin^2\frac{\chi}2]}.
\label{cpr3}
\end{equation}
Strongly anharmonic current-phase relation shows up in \eqref{cpr3}
for $g_\ell^2\gg g_\delta^2\gg 1$, while in the case $g_\ell \ll
g_{\delta}$, $g_\delta^2\gg 1$ the first harmonic dominates the
current and the result \eqref{cpr3} agrees with  \eqref{jtun1}.

The critical current $j_{\text{c}}$, following from \eqref{cpr3}, is
\begin{equation}
j_{\text{c}}=\dfrac{3\sqrt{3}g_\ell j_{\text{dp}}}{4g_\delta(g_\delta
+2g_\ell)}.
\label{jc1}
\end{equation}
It is small $j_{\text{c}} \ll j_{\text{dp}}$ at arbitrary $g_{\ell}$
due to an intense interfacial pair breaking $g_{\delta}^2\gg 1$. At
the sufficiently large $g_\ell\gg g_\delta$ the quantity
$j_{\text{c}}$ in \eqref{jc1}, taken at $g_\delta^2\gg 1$, approaches
the limiting value $(3\sqrt{3}\big/8 g_\delta)j_{\text{dp}}$.

The range of variations of parameters $g_\ell$ and $g_\delta$ is
generally quite wide and includes large values that do not allow to
confine the analysis to the first order terms in $g_\ell$ or
$g_\delta$. Within the GL theory, large positive $g_\delta$ and
$g_\ell$ result in a strong local suppression of the order parameter
$f_0$ at the interface as compared to its bulk value. As seen from
\eqref{bcss9s} and \eqref{cpr3}, under the condition $g_{\delta}^2\gg
1$
\be
f_0^2=\frac{1}{2[g_\delta^2+4(g_\delta+g_\ell)g_\ell
\sin^2\frac{\chi}2]}.
\label{f01}
\ee
The relation $f_0^2\approx 1/(2g_\delta^2)\ll 1$ follows from
\eqref{f01} when $g_\delta^2\gg g_\ell^2$. In the
opposite case $1\ll g_\delta^2\ll g_\ell^2$ a pronounced phase
dependence of the quantity $f_0$ shows up. For small phase
differences $|\chi|\ll g_\delta/g_\ell$ the relation does not change
$f_0^2\approx 1/(2g_\delta^2)\ll 1$, while for $g_\delta/g_\ell\ll
|\chi|\alt\pi$ a stronger suppression takes place $f_0^2\approx 1/(8
g_\ell^2)\ll 1/(2g_\delta^2)\ll 1$. The phase dependent
suppression of $f_0$ is associated not only with the contribution of
the Josephson coupling to the boundary conditions \eqref{bcss99s}, but
also with the current depairing, which is locally enhanced near the
pair breaking interface. Far inside the superconductors the depairing
is small since $j_{\text{c}}\ll j_{\text{dp}}$.

Substituting \eqref{f01} in the boundary conditions \eqref{bcss99s},
one gets $\bigl|\frac{df}{d\overline{x}}\bigr|_{\pm}\sim 1$, since
the interface value of the order parameter $f_0$ is suppressed to
such a degree that the order of magnitude of the right-hand side in
\eqref{bcss99s} retains unchanged. This signifies that in weak links ($j_c\ll
j_{\text{dp}}$) the order parameter always varies on the scale $\xi$,
even if $g_\ell^2\gg 1$.

The phase dependent decrease of $f_0^2$ with $g_\ell$ at $g_\ell\gg
g_\delta$ results in the corresponding decrease of the Josephson
current in an important region of the phase difference, in spite of
the factor $g_\ell$ in \eqref{bcss9s}. The strong suppression of the
order parameter at the junction interface is of key importance in
this paper.

\subsection{Defining relation for $l_j(\Phi)$ in anharmonic junctions}

Within the local Josephson electrodynamics, which
presupposes the condition $l_{\text{j}}\gg \lambda_{\text{L}}$, the
equations for spatial profiles of the quantities along the interface
at $x=0$ meet the standard form
\begin{equation}
\dfrac{dH(y)}{dy}=\dfrac{4\pi}{c}j[\chi(y)], \qquad
\dfrac{d\chi(y)}{dy}=\dfrac{2\pi d}{\Phi_0}H(y),
\label{MaxJos1}
\end{equation}
for the given geometry of the junction and the magnetic field.
Here $\Phi_0={\pi\hbar c}/{|e|}$ is the superconductor flux quantum
and $d=2\lambda_{\text{L}}$, where a small interlayer thickness is
neglected.

The equation for a spatially dependent phase difference follows
directly from \eqref{MaxJos1}:
\begin{equation}
\dfrac{d^2\chi[(y)]}{dy^2}-\,\dfrac{16\pi^2\lambda_{\text{L}}}{c\Phi_0}
j[\chi(y)]=0.
\label{chieq1s}
\end{equation}
Substituting $j[\chi(y)]=j_{\text{c}}\sin[\chi(y)]$ in the right hand
side of Eq.~\eqref{chieq1s}, one gets a well-known one-dimensional
sine-Gordon equation (1) describing the static magnetic properties of
the harmonic junctions.~\cite{Ferrell1963,Josephson1965,Scalapino1967,%
Barone1982}

Under the condition $l_{\text{j}}\alt\lambda_{\text{L}}$ an interplay
of the Meissner and the Josephson screenings takes place, and
Eq.~\eqref{chieq1s} should be modified to incorporate the
corresponding nonlocal effects. The generalized equation has been
obtained for the phase difference in an isolated Josephson vortex far
inside the junctions in question.~\cite{AGurevich1992,Silin1992} It
takes the form
\begin{equation}
\dfrac{c\Phi_0}{16\pi^3\lambda_{\text{L}}^2}\int_{-\infty}^{\infty}du
K_0\left(\frac{|y-u|}{\lambda_{\text{L}}}\right)\dfrac{d^2
\chi(u)}{d u^2}=j[\chi(y)].
\label{chieqnl1}
\end{equation}
While the characteristic scale of the phase difference is
$l_{\text{jv}}$, the Macdonald function $K_0$ in the integrand in
\eqref{chieqnl1} varies on the scale $\lambda_{\text{L}}$. On account
of the relation $2\int_0^\infty K_0(t)dt=\pi$, equation
\eqref{chieqnl1} reduces to \eqref{chieq1s} provided
$\lambda_{\text{J}}\gg\lambda_{\text{L}}$. In a strongly nonlocal
regime $\lambda_{\text{J}}\ll\lambda_{\text{L}}$ the solution of
\eqref{chieqnl1} has been obtained in Ref.~\onlinecite{AGurevich1992}
for the junctions with a harmonic current-phase relation. In
particular, the supercurrent in the Josephson vortex was shown to be
localized mostly over a characteristic length
$\frac{\lambda_{\text{J}}^2}{\lambda_{\text{L}}}\ll
\lambda_{\text{J}}$, with a decaying power law tail on this scale.

In the absence of the magnetic field, the phase dependent
thermodynamic potential per unit area ${\mathit\varOmega}_{S0}(\chi)$
and the supercurrent density $j(\chi)$ are related to each other as
\begin{equation}
j(\chi)=\dfrac{2|e|}{\hbar}\dfrac{d}{d\chi}{\mathit\varOmega}_{S0}(\chi)\,.
\label{jvarOmega1}
\end{equation}
According to the Josephson electrodynamics, the spatial dependence of
the supercurrent arises due to variations of $\chi(y)$ with no
change to $j(\chi)$. A macroscopic scale of the junction penetration
depth allows one to consider, within the local theory, the spatially
dependent quantities ${\mathit\varOmega}_{S0}[\chi(y)]$ and
$j[\chi(y)]$ to be locally related by \eqref{jvarOmega1}, at a given
$y$, as if the phase difference were constant along the interface.

The first integral of Eq.~\eqref{chieq1s} for the anharmonic Josephson
junctions can be obtained after a substitution of the phase-dependent
thermodynamic potential \eqref{jvarOmega1} for the current in
Eq.~\eqref{chieq1s}, and a multiplication of all terms by
$d\chi(y)/dy$. Since the magnetic field is assumed to be fully
screened deep inside the junction plane, one obtains from here making
use of the second equation in \eqref{MaxJos1}
\begin{equation}
H^2(y)=\dfrac{4\pi}{\lambda_{\text{L}}}
\Bigl({\mathit\varOmega}_{S0}[\chi(y)]
-{\mathit\varOmega}_{S0}(0)\Bigr)\!=
\dfrac{2\Phi_0}{c\lambda_{\text{L}}}\int_0^{\chi(y)}
\!\!\!\! j(\chi)d\chi ,
\label{int13pp}
\end{equation}
where ${\mathit\varOmega}_{S0}[\chi_{\infty}=0]\equiv {\mathit
\varOmega}_{S0}(0)$.

After integrating Eqs.~\eqref{MaxJos1} along the y-axis, under the
given conditions one also gets
\begin{equation}
H\left(0\right)=-\dfrac{4\pi}{cL_z}I, \qquad
\chi\left(0\right)=-\dfrac{2\pi}{\Phi_0}\Phi\, ,
\label{chiPhi1pp}
\end{equation}
where $I$ and $\Phi=2\lambda_{\text{L}}\int_0^{\infty} H(y)dy=
2\lambda_{\text{L}}l_{\text{j}} H(0)$ are the total current and
the magnetic flux through the junction.

The first relation in \eqref{chiPhi1pp} and (7) allow one to
identify the total current through the harmonic junction of
Ferrell and Prange, as a function of the magnetic flux:
\begin{equation}
I=-\dfrac{cL_z\Phi_0}{8\pi^2\lambda_{\text{J}}\lambda_{\text{L}}}
\sin\left(\dfrac{\pi\Phi}{\Phi_0}\right).
\end{equation}

Having in mind the current-phase relations \eqref{jgl3} -
\eqref{cpr3}, one considers $0$-junctions ($g_\ell>0$) with
$2\pi$-periodic anharmonic phase dependent thermodynamic potential,
which has only one minimum and one maximum per period. One notes that
this condition excludes from the consideration the $\varphi$-junctions
with several different solutions for an isolated Josephson
vortex.~\cite{Buzdin2003,Goldobin2007,Goldobin2012} The
$2\pi$-periodicity of ${\mathit \varOmega}_{S0}(\chi)$ and $j(\chi)$
allows a global shift of $\chi$ by $2\pi n$ ($n=\pm1, \pm2, \dots$).
The values $\chi=2\pi n$ (and, in particular, $\chi=0$)
correspond to the minima of ${\mathit\varOmega}_{S0}(\chi)$ and to
the vanishing supercurrent. As seen from \eqref{int13pp} and
\eqref{chiPhi1pp}, by fixing the phase difference $\chi(0)$ at the
junction edge, one simultaneously specifies the total magnetic flux
$\Phi$ penetrating through the junction, the magnetic field $H(0)$ at
the junction edge and the total Josephson current $I$. Thus the
thermodynamic potential ${\mathit\varOmega}_{S0}[\chi(0)]$ can also
be considered, for example, as a function of the magnetic field
$H(0)$, or the magnetic flux $\Phi$. Eqs.~\eqref{int13pp} and
\eqref{chiPhi1pp} also ensure, for given current-phase relations,
that the maxima of $|H(0)|$,\,${\mathit\varOmega}_{S0}[\chi(0)]$ and
$|I|$ take place at the phase differences $\chi(0)=2\pi\left(n+
\frac12\right)$. The magnetic flux at these phase differences takes
half-integral values of the flux quantum, and the Josephson current
density $j[\chi(0)]$ vanishes at the junction edge.

The magnetic penetration depth $l_{\text{j}}(\Phi)$ as a function of
the magnetic flux through the junction with an anharmonic
current-phase relation, follows from the relation $\Phi=2
\lambda_{\text{L}} l_{\text{j}} H(0)$, where one should take into account
\eqref{int13pp} and \eqref{chiPhi1pp}:
\begin{equation}
l_{\text{j}}^{-1}(\Phi)=\left\{\frac{16\pi \lambda_{\text{L}}}{\Phi^2}\!
\left[{\mathit\varOmega}_{S0}\Bigl(\frac{2\pi\Phi}{\Phi_0}\Bigr)
-{\mathit\varOmega}_{S0}\bigl(0\bigr)\right]\right\}^{1/2}\!\!\!\! .
\label{deflambdaj2s}
\end{equation}
Here ${\mathit\varOmega}_{S0}(\chi)$ is assumed to be an even
function of $\chi$.

The defining relation (9) for the junction penetration depth follows
from \eqref{deflambdaj2s} and \eqref{int13pp}.

For the junctions with a pronounced interfacial pair breaking
$g_\delta^2\gg 1$ one gets from \eqref{jvarOmega1} and \eqref{cpr3}
\begin{multline}
{\mathit\varOmega}_{S0}[\chi(y)]-{\mathit\varOmega}_{S0}(0)=
\dfrac{\hbar j_{\text{dp}}}{2|e|}\int_0^{\chi(y)}j(\chi)d\chi=\\ =
\dfrac{3\sqrt{3}\hbar j_{\text{dp}}}{16|e| (g_\delta+g_\ell)}
\ln\left[1+\dfrac{4g_\ell(g_\delta+g_\ell)}{g_\delta^2}\sin^2\dfrac{\chi(y)}{2}\right].
\label{omegagdeltagg1}
\end{multline}

Eq.~(10) follows from \eqref{deflambdaj2s}, \eqref{omegagdeltagg1}
and from the expressions for $j_{\text{dp}}$, $\lambda_{L}$ and $\xi$.

\subsection{Basic equation for $\chi(y)$ in the vortex in anharmonic
junctions}

For describing the spatial dependence of the phase difference in the
Josephson vortex, it is convenient to base it on the equation, which
follows from Eq.~\eqref{int13pp} after the derivative of the phase
difference is substituted for the magnetic field using \eqref{MaxJos1}.
Introducing the dimensionless coordinate $\tilde{y}=
\frac{y}{l_{\text{jv}}}$, where $l_{\text{jv}}=
l_{\text{j}}(\frac{\Phi_0}{2})$ and $l_{\text{j}}$ is defined in
\eqref{deflambdaj2s}, one gets
\begin{equation}
\left(\dfrac{d\chi}{d\tilde{y}}\right)^2=\pi^2
\dfrac{{\mathit\varOmega}_{S0}[\chi(y)]
-{\mathit\varOmega}_{S0}(0)}{{\mathit\varOmega}_{S0}(\pi)
-{\mathit\varOmega}_{S0}(0)}.
\label{derchi1}
\end{equation}

Eq.~(13) follows from \eqref{derchi1} and \eqref{omegagdeltagg1}.

\subsection{Basic expression for $H_{\text{jc}1}$ in anharmonic junctions}

The lower critical field of the junction is known to satisfy the
relation $H_{\text{jc}1}=4\pi {\mathit\varOmega}_{l}\bigl/\Phi_0$,
where ${\mathit\varOmega}_{l}$ is the thermodynamic potential of
the Josephson vortex per unit length. The quantity
${\mathit\varOmega}_{l}$ can be written as
\begin{equation}
{\mathit\varOmega}_{l}=\int\limits_{0}^{\infty}
\left\{\dfrac{H^2(y)}{8\pi}2\lambda_{\text{L}}+
\Bigl[{\mathit\varOmega}_{S0}[\chi(y)]-{\mathit\varOmega}_{S0}(0)\Bigr]
\right\}dy.
\label{funcen1p}
\end{equation}
It includes both the magnetic field energy and the junction term
${\mathit\varOmega}_{S0}[\chi(y)]$ in the absence of the magnetic
field, taken for the phase difference $\chi(y)$. The spatial
dependence of $\chi(y)$ is described by a solution of
Eq.~\eqref{derchi1} for an isolated Josephson vortex and thereby it
takes into account the self field effects. The relation
\eqref{int13pp} shows that the two terms in \eqref{funcen1p} are
equal to each other. Hence,
\begin{equation}
H_{\text{jc}1}=\dfrac{8\pi l_{jv}}{\Phi_0}\int\limits_{0}^{\infty}
\Bigl[{\mathit\varOmega}_{S0}[\chi(y)]-{\mathit\varOmega}_{S0}(0)\Bigr]d\tilde{y}.
\label{funcen1pp}
\end{equation}

Eq.~(14) for $H_{\text{jc}1}$ follows from \eqref{funcen1pp}, \eqref{omegagdeltagg1}
and (11) after taking into account the expressions for $j_{\text{dp}}$,
$\lambda_{L}$, $\xi$ and $\lambda_J$.

\end{document}